\documentclass[amsmath,amssymb,aps,12pt,prd,letterpaper,titlepage,onecolumn,floatfix,superscriptaddress,nofootinbib,preprintnumbers]{revtex4-2}

\usepackage{graphicx}
\usepackage{bm}
\usepackage{epstopdf}
\usepackage{feynmp-auto}
\usepackage{xcolor}
\usepackage[utf8]{inputenc}
\usepackage{slashed}
\usepackage{ragged2e}

\usepackage[sort&compress]{natbib}	 
	\setcitestyle{square,numbers,comma}
\usepackage[colorlinks=true,urlcolor=blue,linkcolor=black,citecolor=blue]{hyperref}

\usepackage{fontawesome5}
\definecolor{orcidlogocol}{rgb}{0.65, 0.807, 0.223}
\newcommand{\orcid}[1]{\,\href{https://orcid.org/#1}{\textcolor{orcidlogocol}{\footnotesize\faOrcid}}}

\begin{document}

\preprint{\texttt{FERMILAB-PUB-24-0325-SQMS-TD}}

\title{A Precision Gyroscope from the Helicity of Light}
\date{\today}

\author{Michael A.~Fedderke\orcid{0000-0002-1319-1622}}
\email{mfedderke@perimeterinstitute.ca}

\affiliation{Perimeter Institute for Theoretical Physics, Waterloo, Ontario, N2L 2Y5, Canada}
\affiliation{The William H. Miller III Department of Physics \& Astronomy, The Johns Hopkins University, Baltimore, Maryland  21218, USA}

\author{Roni Harnik\orcid{0000-0001-7293-7175}}
\affiliation{Fermi National Accelerator Laboratory, Batavia, Illinois 60510, USA}

\author{David E.~Kaplan\orcid{0000-0001-8175-4506}}
\email{david.kaplan@jhu.edu}
\affiliation{The William H. Miller III Department of Physics \& Astronomy, The Johns Hopkins University, Baltimore, Maryland  21218, USA}

\author{Sam Posen\orcid{0000-0002-6499-306X}}
\affiliation{Fermi National Accelerator Laboratory, Batavia, Illinois 60510, USA}

\author{Surjeet Rajendran\orcid{0000-0001-9915-3573}}
\email{srajend4@jhu.edu}
\affiliation{The William H. Miller III Department of Physics \& Astronomy, The Johns Hopkins University, Baltimore, Maryland  21218, USA}

\author{Francesco Serra\orcid{0000-0002-7962-3555}}
\email{fserra2@jh.edu}
\affiliation{The William H. Miller III Department of Physics \& Astronomy, The Johns Hopkins University, Baltimore, Maryland  21218, USA}

\author{Vyacheslav P.~Yakovlev\orcid{0000-0003-3969-9421}\,}
\affiliation{Fermi National Accelerator Laboratory, Batavia, Illinois 60510, USA}

\begin{abstract}
We describe a gyroscope that measures rotation based on the effects of the rotation on the polarization of light. Rotation induces a differential phase shift in the propagation of left- and right-circularly polarized light and this phase shift can be measured in suitably designed interferometric setups. The signal in this setup is independent of the frequency of light, unlike various sources of noise such as vibrations, which cause phase shifts that depend on the frequency. Such vibrations are the practical limit on the sensitivity of conventional Sagnac-style optical interferometers that are typically used as gyroscopes. In the proposed setup, one can potentially mitigate this source of noise by simultaneously using two (or more) sources of light that have different frequencies. The signal in this setup scales with the total storage time of the light. Due to its frequency independence, it is thus most optimal to measure the signal using superconducting radio-frequency systems where the high finesse of the available cavities enables considerably longer storage times than is possible in an optical setup. 
\end{abstract}

\maketitle

\section{Introduction}

Precision gyroscopes have a number of important uses, ranging from inertial navigation to fundamental physics applications such as tests of General Relativity. A number of precision techniques have been developed to measure rotations. This includes the use of the Sagnac effect in optical~\cite{GEStedman_1997, LEFEVRE2014851, Culshaw_2006} and atomic interferometers~\cite{doi:10.1126/sciadv.aau7948}, measurements of nuclear spin precession~\cite{PhysRevLett.120.033401, PhysRevLett.95.230801} and gyroscopy based on the Josephson effect in superfluid helium SHeQUIDs~\cite{Packard_2014}. Given the broad range of applications for precision gyroscopy, it is interesting to develop new techniques that can potentially achieve greater sensitivity. In this paper, we propose a new method to measure rotation by looking for its effects on the polarization of light. 

The basic physical principle behind this measurement scheme is as follows. Rotation changes the dispersion relation of circularly polarized light (see, e.g., \cite{PhysRevD.41.1231,PhysRevD.43.3789,HARARI199267,PhysRevLett.81.3067,DeRocco:2018jwe,Liu:2018icu,Fedderke:2019ajk}), leading to a differential  shift in the dispersion relation for left- and right-circularly polarized light. This shift leads to a relative phase between suitably stored right- and left-circularly polarized light, which can be measured precisely using interferometry. The key advantage of this method of measurement, as we will show below, is that the phase shift is independent of the frequency of light: it depends only on the rotation rate and the time for which the light is stored. This is unlike the case of the Sagnac effect that is typically used to measure rotation using light, where the rotation manifests as a length change. In that case, the phase shift is directly proportional to the frequency of light.  A key limitation of gyroscopes based on the Sagnac effect is vibration of the mirrors used in the setup; such vibrations cause relative length changes in the apparatus which lead to phase shifts that are also directly proportional to the frequency of light, preventing the system from distinguishing between rotations and vibrations. However, in the measurement scheme proposed in this paper, the phase shift caused by polarization is independent of the frequency of the light whereas phase shifts arising from vibrations will be directly proportional to the frequency. Due to this difference, one can envisage a setup where two frequencies of light are used to measure simultaneously, breaking the degeneracy between the signal from rotation and the noise from vibration. That is, because rotations and vibrations cause phase shifts that scale differently with frequency, the phase-shift difference between the two frequencies can be used to measure and subtract out vibrations, giving access to the signal from rotations that is common to both frequencies. 

The independence of the signal on the frequency of light also implies that the signal is the same whether we use optical or radio frequencies (RF) to measure the signal. Since the signal also scales with the storage time of the light, it is advantageous to use radio-frequency light for this measurement since these frequencies can be stored for considerably longer periods of time (for example, in superconducting cavities) than optical light. We will thus consider the use of suitably designed RF cavities to detect this effect in the rest of this paper, although the same technique can be used at optical frequencies, but with reduced overall sensitivity. 

The rest of this paper is organized as follows. In Sec.~\ref{sec:signal}, we compute the effects of rotation on left- and right-circularly polarized light. Following this, in Sec.~\ref{sec:setup} we propose a measurement scheme to detect this effect in a suitably designed Fabry--P{\'e}rot cavity. We estimate the fundamental sensitivity of such a scheme in Sec.~\ref{sec:sensitivity} and compare the reach of this scheme to that of the conventional Sagnac scheme. We also comment on various systematic effects that may prevent us from achieving this fundamental limit and suggest some potential ways to mitigate these effects. 
Following this, we conclude in Sec.~\ref{sec:conclusions}. 
In Appendix \ref{app:rotatingframe} we give an explicit derivation of our solution for a plane wave in the rotating frame.
In Appendix \ref{app:generalRotations} we give more general discussion of solutions for light propagating in rotating cavity setups other than those discussed in the main text. 
The analysis in the main text of this paper is performed in a rotating frame; we describe the analysis in the inertial frame in Appendix \ref{app:Inertial}.

\section{The Signal}
\label{sec:signal}
In this section, we compute the effect of rotation on the polarization of light. We begin with a simple situation: assume that there is a system rotating around the $z$ axis with a rotation rate $\Omega$. Imagine that from the origin, we send left- or right-circularly polarized light along the $z$ axis. How is the polarization affected by the rotation? 

Let us analyze this system first in the rotating frame. In this frame, the metric that describes the system is: 
\begin{equation}
ds^2 = \left[ - 1 + \Omega^2 \left(x^2 + y^2\right)\right] dt^2 - 2 \,\Omega \, y\, dt dx + 2\, \Omega \, x \, dt dy + dx^2 + dy^2 + dz^2\: . \label{eq:metric}
\end{equation}

Working in the Weyl gauge $A_0 = 0$, consider the propagation of two electromagnetic waves with vector potentials: 
\begin{equation}
(A_{\pm})_\mu = \mathcal{A}_\pm^0 e^{i \left(\omega t - k_{\pm} z\right)} \big[ 0,  1,  \pm \,i,A_{z}\left(t, x, y, z\right) \!\big]\:, \label{eq:simplifiedSoln}
\end{equation}
where $\mathcal{A}_{\pm}$ correspond, respectively, to right- ($-$) and left- ($+$) circularly polarized light.%
\footnote{\label{ftnt:PolConvention}%
    Given our sign conventions, the left/right definitions for handedness adopted here match those seen by the emitter.
} %
Here, $A_z = \pm i \Omega (x\pm i y)$, which vanishes in the limit $x \rightarrow 0$, $y \rightarrow 0$ (appropriate when the transverse size of the light beam is small compared to $1/\Omega$).
Waves of the form at \eqref{eq:simplifiedSoln} are solutions to the vacuum Maxwell equations that have the following dispersion relations:
\begin{equation}
k_{\pm} = \omega \mp \Omega\:.
\label{Eqn:Dispersion}
\end{equation}
We give a derivation of this result in Appendix \ref{app:rotatingframe}.
This shows that right- and left-circularly polarized light will travel with different wavenumbers $k_{\pm}$. If they travel for a distance $L$, the differential phase between the right- and left-circularly polarized light will be $\Omega L$, independent of the frequency $\omega$ of the light. 
We discuss more general orientations for the rotation with respect to the propagation direction of the light in Appendix \ref{app:generalRotations}.

How can we understand this result in an inertial frame? In this frame, the propagation of the light is unaffected by the rotation since the light simply moves in Minkowski space. However, the source of the polarized light is rotating in this frame. This leads to a continually increasing relative phase between the propagating light and the polarization axis of the source that produces the light. 

The signal is thus similar to that which arises in nuclear spin gyroscopes. In these, the nuclear spins act as an inertial reference, with their orientation unaffected by rotation. However, rotation causes the sensing apparatus to rotate, resulting in a relative rotation between the spin and the measuring apparatus. Similarly, in the case of this signal, the polarized states of the light act as inertial references. The signal then arises due to the relative rotation between this inertial reference and the rotating apparatus used to produce/detect the polarized light.

\section{The Setup}
\label{sec:setup}

\begin{figure}[h]
\includegraphics[width=1.0\textwidth]{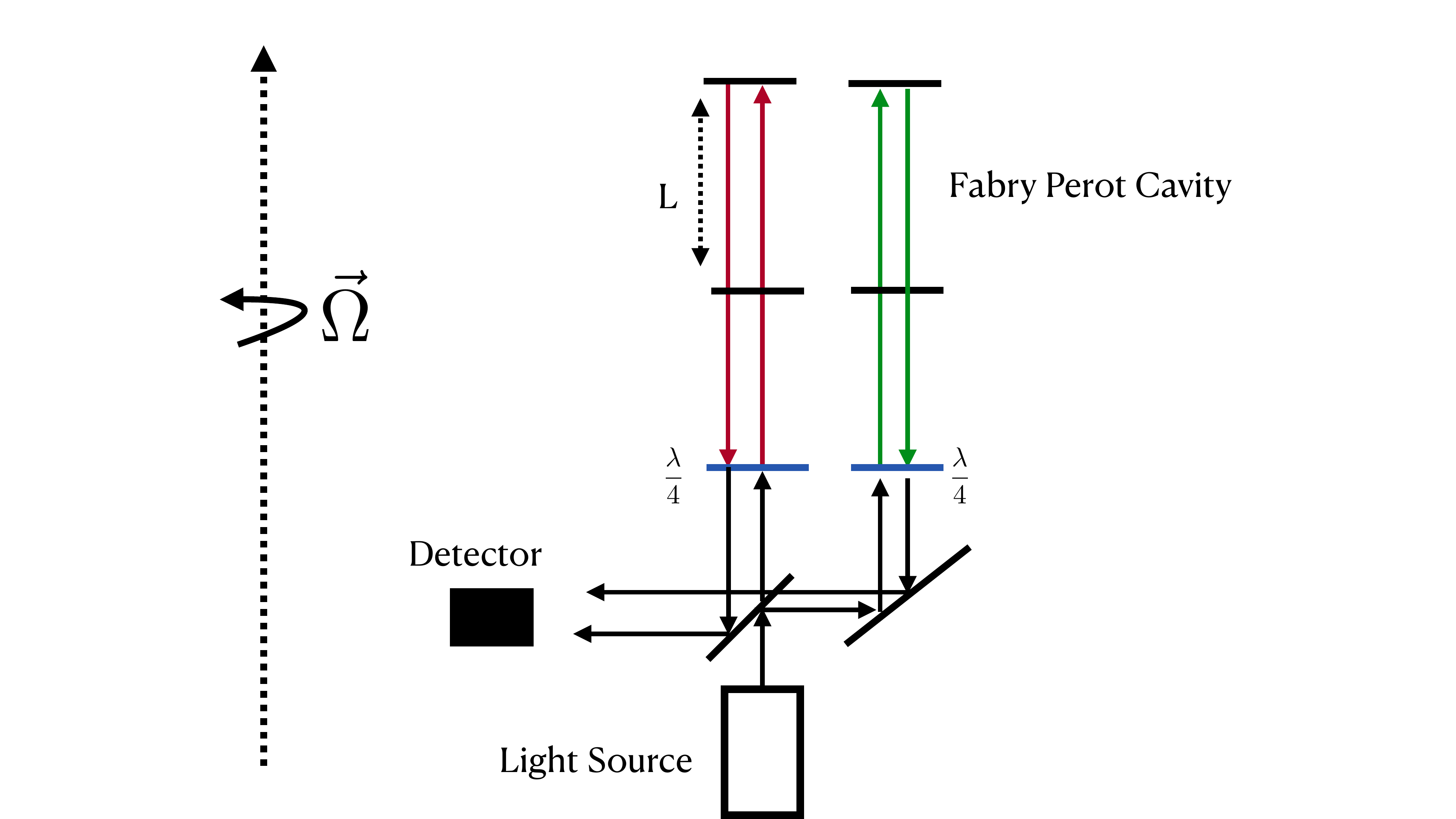}
\caption{The scheme to measure rotation. Linearly polarized light is sent through a suitable polarization apparatus which splits the beam into left- (red) and right- (green) circularly polarized beams. These are stored in two Fabry--P{\'e}rot cavities that are aligned with the axis of rotation $\vec{\Omega}$. The differential phase between these two beams is free of phase and frequency noise from the light source, but retains the rotation signal.}
\label{Fig:Setup}
\end{figure}

The optimal way to measure the shift \eqref{Eqn:Dispersion} in the wavenumber caused by the rotation is to convert it into a phase. To maximize this phase, it is advantageous to expose the light to the rotation for as long as possible. In order to achieve this in a compact device, the light has to be held in a cavity of some kind. Further, we also need a measurement scheme that significantly suppresses dominant sources of noise such as frequency and phase noise in the light, as well as systematics that may arise from vibrations in the cavity. Motivated by these considerations, we propose the setup described in Fig.~\ref{Fig:Setup}. 

In this setup, we have two Fabry--P{\'e}rot cavities, each of length $L$ and finesse $\mathcal{F}$. The cavities are aligned along the direction of rotation  $\vec{\Omega}$. We produce linearly polarized light from a suitable source. This light is split and one part of it is sent to a suitable optical element [i.e., a quarter-wave ($\lambda/4$) plate] which converts it to left-circularly polarized light while the other part is sent to a different quarter-wave plate that converts that part to right-circularly polarized light. The two beams are sent to two different Fabry--P{\'e}rot cavities where they are stored for a time $\sim \mathcal{F} L$. After this time, the beams are sent back through the quarter-wave plates and recombined. The interfered beam is detected at the detector. 

In this scheme, the wavenumbers of the left- and right-circularly polarized light experience different shifts. This difference manifests as a relative phase shift between these two beams when they are re-interfered.  We want the signal in this setup to scale with $\mathcal{F} L$, the total time for which the light is held in the cavity. If the mirrors of the cavity are made with conventional reflecting surfaces, the overall phase of the electromagnetic vector potential $\left(A_{\pm}\right)_{\mu}$ changes by $\pi$ upon reflection. But, the reflection does {\it not} cause a  {\it relative} phase between the $x$ and $y$ components of $\left(A_{\pm}\right)_{\mu}$. It can then be shown that for the reflected wave also the wavenumbers are $k_{\pm} = \omega \mp \Omega$. Thus, the phase accumulated by the light will continually {\it add}. In other words, while left- (right-) circularly polarized light becomes right- (left-) circularly polarized upon reflection, since the direction of the light has also changed, the effect of the rotation on the wavenumber remains the same.%
\footnote{The situation is thus similar to the accumulation of phase for light stored in a cavity in the presence of a static axion gradient (cf.~\cite{Fedderke:2023dwj}), as opposed to a long-period time-dependent dark-matter axion field for which the phase accumulation cancels out when employing conventional mirrors in the cavity (cf.~\cite{DeRocco:2018jwe}).} %
Another way to understand the continual addition of phase regardless of the direction of travel of the light in the cavity is that the flip of left- to right-handedness (and vice versa) of the light upon reflection off the mirrors is an \emph{helicity} flip that occurs because the linear momentum of the light changes sign but the angular momentum does not.
Because the relative orientation of the angular momentum of the light and that of the rotation of the cavities is however unchanged after the bounce off the mirror, the phase shift to the light induced by the rotation continues to add for trips in either direction in the cavity.

The phase shift between the two arms of the interferometer is (see, e.g., \cite{Fedderke:2023dwj}):%
\begin{equation}
\Delta \Phi \sim \frac{4\mathcal{F}}{\pi} \Omega  L \:.
\label{Eqn:PhaseShift}
\end{equation}
In general, this signal can be either larger or smaller by a factor of $\mathcal{O}(1)$ depending on choices made for the relative cavity orientations and on the orientation of the rotation axis with respect to the cavity orientation.
Notice that since the phase shift $\Delta \Phi$ is a differential phase, it is free of frequency and phase noise inherent to the light source. Relative motions  of  the mirrors of the cavity will also cause uncancelled phase shifts in this setup. But, crucially, the phase shift caused by a relative length shift $\delta L$ is $\omega \, \delta L$, where $\omega$ is the frequency of the light. This is frequency dependent, unlike \eqref{Eqn:PhaseShift} which is frequency independent. Since low-frequency vibrations are a limiting source of noise for conventional Sagnac gyroscopes, we now see the potential advantage of the proposed measurement scheme: one can operate the entirety of the above setup with two different frequencies of light simultaneously. Since the phase shifts from the rotational signal and vibrational noise scale differently with the frequency of the light, we can use the difference between the phase shifts at the two frequencies as an effective measurement of the vibrations in the system, allowing us to subtract it off and gain access to the common phase-shift between the two frequencies that encodes the rotation rate $\Omega$. 

The setup described above can be implemented with a variety of light sources, whether optical or RF. However, there is a distinct advantage to using RF sources instead of optical light. The signal \eqref{Eqn:PhaseShift} is independent of the frequency $\omega$ of the light and only depends on the storage time $\mathcal{F} L$ of the light. Due to the existence of ultra-high-finesse superconducting RF cavities, RF light can be stored for considerably longer periods than optical light. From this perspective, it is thus advantageous to realize this setup using RF rather than optical sources. The disadvantage of the RF source though is that the longer wavelength of the RF light requires larger mirrors for the cavities to combat diffractive losses and thus gyroscopes using RF light may not be useful in applications where the compactness of the device is a significant design constraint.

\section{Sensitivity}
\label{sec:sensitivity}

The fundamental limit on the sensitivity of this setup is set by the shot noise limit on the resolution of the phase shift \eqref{Eqn:PhaseShift}. This yields (e.g., \cite{Fedderke:2023dwj}):%
\begin{equation}
\Omega \sim 10^{-11} \frac{\text{Hz}}{\sqrt{\text{Hz}}} \, \left(\sqrt{\frac{10^{9}}{\mathcal{F}}} \right) \, \left(\frac{1\,\text{m}}{L}\right) \sqrt{\frac{1\,\text{MW}}{P_{\rm max}}} \sqrt{\frac{\omega}{2 \pi  \times 10\,\text{GHz}}} \label{eq:shotNoise}
\end{equation}
for a device operating with a circulating power $P_{\rm max}$. We have used fiducial parameters suitable for superconducting RF cavities in this estimate. Note that the required input power to the cavity is only $\sim  P_{\rm max} / \mathcal{F} \sim 1\,\text{mW} \times (P_{\rm max} / 1\,\text{MW}) \times (10^9/\mathcal{F})$.
In the estimate at \eqref{eq:shotNoise}, we took a finesse of $\mathcal{F} \sim 10^9$.
For reasons related to RF cavity control at such high finesse, this may be an aggressive assumption; see, e.g., the extended discussion of this point in a related context in \cite{Fedderke:2023dwj}. 
A more conservative assumption would be $\mathcal{F} \sim 10^7$, for which cavity control requirements are significantly relaxed; this would however only degrade our sensitivity estimate at \eqref{eq:shotNoise} by one order of magnitude, and the input power required would still only be $\sim 0.1\,\text{W}$.

The theoretical fundamental sensitivity of this device is thus greater than the demonstrated sensitivities $\sim 10^{-8} \frac{\text{Hz}}{\sqrt{\text{Hz}}}$  of terrestrial gyroscopes \cite{doi:10.1126/sciadv.aau7948} that are constructed with optical or atom interferometers in the Sagnac configuration, or nuclear-spin-based gyroscopes. \\

Given this limit, it is interesting to compare the sensitivity of our proposed approach to the theoretical sensitivities of the Sagnac setup. In the Sagnac configuration, where the arm lengths of the setup are $L$, the phase shift is (see, e.g., \cite{Culshaw_2006})
\begin{equation}
\Delta \Phi_{S} \sim \mathcal{F} \, \omega \,  \Omega \,  L^2\:,
\label{Eqn:Sagnac}
\end{equation}
assuming for the purposes of comparison that the Sagnac configuration is operating such that the light completes $N \sim \mathcal{F}$ loops around the Sagnac ring before being read out. Comparing \eqref{Eqn:Sagnac} to \eqref{Eqn:PhaseShift}, we see that the conventional Sagnac phase shift is parametrically a factor of $\omega L$ larger than the polarization phase shift that we have considered in this paper. For optical light, $\omega L \gg 1$ is typical for a macroscopic cavity; for RF, $\omega L$ is typically not as large, leading to a smaller parametric enhancement for an RF Sagnac configuration.

A naive reading of this result would suggest that our proposed approach is therefore inferior. However, this result must be interpreted with nuance and care: while it is true that a Sagnac interferometer has a fundamental shot-noise limited sensitivity that, for the same assumed parameters, is parametrically enhanced as compared to the approach we propose, most practical Sagnac interferometers are limited not by their shot-noise floor, but instead by vibrational noise sources that cannot \emph{even in principle} be distinguished by the signal of rotation. To understand why this is the case, consider that in a typical Sagnac configuration, light propagates in two different directions along a closed loop. Rotation causes one of these paths to be longer than the other, resulting in a phase shift signal that is proportional to the frequency $\omega$ of the light. In the absence of rotation, suppose one of the optical elements in the Sagnac interferometer instead experiences a low-frequency vibrational motion. Since the light rays traverse the Sagnac loop along different directions, they experience the position of this vibrating optical element at slightly different times. This results in an uncommon length difference between the two light paths, leading to a phase shift that is \emph{also proportional to the frequency of the light, $\omega$}. In a Sagnac setup therefore, there is no way, even in principle, to distinguish a signal of rotation from a vibration of the optical elements; vibrations therefore constitute an irreducible noise floor.

By contrast, our proposed approach makes it possible to distinguish vibrational noise of the optical elements from the rotation signal. We envisage the simultaneous use of two different (cavity-resonant) frequencies of light in making our measurement. If one considers all light arriving at the interferometric readout at a given time then (assuming the finesse is flat with frequency) all its frequency components will have always followed the same optical path in the apparatus, and will thus have encountered the same stochastically vibrating optical elements everywhere in the optical path at the same times. The implication is that all frequencies will experience a common (although stochastic over time) length fluctuation of the two cavities during the time the light resides in the apparatus. This common length fluctuation will however give rise to different, although proportional, phase shifts at each light frequency. However, the rotation signal \eqref{Eqn:PhaseShift} in this setup is independent of the light frequency. It is therefore in principle possible to distinguish the phase noise arising from vibrations of optical elements from a true rotation signal in our proposed measurement: the vibrational signal shows up in the differential phase shift between the two frequencies, and can be subtracted off to expose the rotationally induced phase that is common to both (or all) frequencies.
This ability of our proposed technique to ameliorate vibrational noise could thus enable it to achieve greater sensitivity in practice than is achievable with the Sagnac configuration.

We now comment on an important systematic issue that may prevent the proposed gyroscope from achieving the fundamental sensitivity \eqref{Eqn:PhaseShift}; namely, potential shifts to the polarization vector of the light whenever the light reflects from the mirror.  This effect would scale with the finesse $\mathcal{F}$, but it is independent of the cavity length $L$. This is unlike the signal from the rotation which scales with $L$. Given this difference, it should, in principle, be possible to mitigate the effects of such a shift. One could operate the gyroscope with various cavity lengths and use the difference in the functional dependence of the signal and this systematic to calibrate this effect out. With such a calibration, the systematic could still affect this measurement if there is a time dependent contribution to this polarization shift, leading to a change in the magnitude of the effect between the calibration and the operation of the device. It is reasonable to expect that in a superconducting RF setup, the cause of such a polarization shift is due to mechanical imperfections in the mirrors themselves and that these imperfections are likely to be static. It is thus reasonable that the time dependence of these shifts is small, but this question would have to be carefully studied under realistic experimental conditions. 

The envisaged operation of this light-based gyroscope would be as a final fine-measurement stage. That is, we presume that this gyroscopic assembly would be mounted on a set of nulling gimbals designed to cancel most of the gross rotational motion of the platform being monitored. It is for this reason that we have analyzed only rotational motion oriented around the axis of the light propagation; large rotational motion around another axis would induce additional phase shifts and phase gradients across cavity mirrors that would complicate the analysis. If fine measurement along multiple axes is required, either a set of orthogonal light gyroscopes could be used, or the additional phase shifts induced by rotation about an axis different from the cavity symmetry axis would need to be understood in more detail.
See also Appendix \ref{app:generalRotations} for further discussion.

Moreover, it is likely that in order to impute the high degree of rotational measurement precision made available by this light gyroscope to a measurement of the platform whose rotational state is desired to be known, either an extremely rigid set of mounts to the platform would be required, or alternatively active metrology and feedback would be needed to measure and control for any relative motion of the gyroscope assembly and the platform due to, e.g., vibrations.

\section{Conclusions}
\label{sec:conclusions}
In this paper, we have outlined a new protocol that measures rotation through its effects on the polarization of light. The independence of the rotation signal on the frequency of light enables this setup to potentially combat systematics arising from vibrational noise that confront typical Sagnac-style interferometric measurements of rotation. Further, the frequency independence of the signal also means that one could measure this signal using high-finesse RF systems where the long storage time of the light can significantly enhance the sensitivity of the apparatus. If successfully developed, the technology developed here may have broad applications in commercial and scientific areas. From the perspective of fundamental physics, it might be interesting to investigate if RF gyroscopes can be used to detect physics that uniquely impacts the spin of light, such as the Lense--Thirring effect of General Relativity or dark-matter candidates such as axions and/or axion-like particles.

\acknowledgments
This work was supported by the U.S.~Department of Energy~(DOE), Office of Science, National Quantum Information Science Research Centers, Superconducting Quantum Materials and Systems Center~(SQMS) under Contract No.~DE-AC02-07CH11359. D.E.K.~and S.R.~are supported in part by the U.S.~National Science Foundation~(NSF) under Grant No.~PHY-1818899.
S.R.~is also supported by the Simons Investigator Grant No.~827042, and by the~DOE under a QuantISED grant for MAGIS. 
D.E.K.~is also supported by the Simons Investigator Grant No.~144924.
Research at Perimeter Institute is supported by the Government of Canada through the Department of Innovation, Science, and Economic Development, and by the Province of Ontario through the Ministry of Colleges and Universities.\\

{\bf Note Added:} After the appearance of the \texttt{arXiv v1} preprint of this work, we became aware of~\cite{MASHHOON1989103}, which discusses the effect of rotation on the polarization of light, including the frequency independence of the signal. But,~\cite{MASHHOON1989103} does not exploit the fact that the frequency independence of the signal can be used to implement this measurement strategy using RF light instead of optical light. The use of RF light considerably enhances the sensitivity of the measurement scheme since it can be stored for much longer time than optical light, enabling the setup to reach sensitivities comparable to a Sagnac configuration that is shot-noise limited. Moreover,~\cite{MASHHOON1989103} also does not exploit the frequency independence of the signal to suppress phase shifts arising from vibrations, one of the major sources of noise in conventional Sagnac gyroscopes.

\appendix

\section{Rotations around the \texorpdfstring{$\hat{z}$}{z} axis}
\label{app:rotatingframe}

In this appendix we derive the results of Eqs.~\eqref{eq:simplifiedSoln} and \eqref{Eqn:Dispersion}.
Let us consider Maxwell's equations in a frame rotating around the $\hat{z}$ axis $\vec{\Omega}=(0,0,\Omega)$. 
We have the following metric:
\begin{align}
    g_{\mu\nu}dx^\mu dx^\nu=[-1+\Omega^2(x^2+y^2)]dt^2+2\Omega dt(x dy-y dx)+dx^2+dy^2+dz^2\:.
\end{align}
Maxwell's equations will be $\nabla_\mu F^{\mu\nu}=0$, where $\nabla_\mu$ is a covariant derivative with respect to $g_{\mu\nu}$.
Looking for plane-wave solutions propagating along $z$, we can pick Weyl gauge and take the following Ansatz for the vector potential:
\begin{align}
    A_{\mu} = \big[ 0, A_x(z,t),A_y(z,t),A_z(x,y,z,t)\big] \:.
\end{align}
As we show here, this Ansatz solves Maxwell's equations in the rotating frame.
Indeed, computing $\nabla_\mu F^{\mu\nu}=0$, we find the following:
\begin{align}\label{eqsorigin}
    \begin{split}
        &\partial_t\partial_z A_z+\Omega\left[\partial_z^2(xA_y-yA_x)+(y\partial_x-x\partial_y)\partial_zA_z\right]=0\:,\\
        &(\partial_t^2-\partial_z^2)A_x+\partial_{x}\partial_z A_z\\&\quad\quad\quad=\Omega(\partial_tA_y+y\partial_t\partial_z A_z)+\Omega^2\left[xy\partial_z(\partial_zA_y-\partial_y A_z)-y^2\partial_z(\partial_zA_x-\partial_x A_z)\right]\:,\\
        &(\partial_t^2-\partial_z^2)A_y+\partial_{y}\partial_z A_z\\&\quad\quad\quad=-\Omega(\partial_tA_x+x\partial_t\partial_z A_z)+\Omega^2\left[xy\partial_z(\partial_zA_x-\partial_x A_z)-x^2\partial_z(\partial_zA_y-\partial_y A_z)\right]\:,\\
        &(\partial_t^2-\partial_x^2-\partial_y^2)A_z-\Omega\partial_t\left[\partial_z(yA_x-xA_y)+2(x\partial_y-y\partial_x)A_z\right]\\&\quad\quad\quad=\Omega^2\left[x(\partial_xA_z-\partial_z A_x)+y(\partial_yA_z-\partial_zA_y)-(x^2\partial_y^2+y^2\partial_x^2-2xy\partial_x\partial_y)A_z\right]\:.
    \end{split}
\end{align}
These equations can be solved with the following choice:
\begin{align}
    A_\mu=e^{i(\omega t-k_\sigma z)}\big[0,1,i\sigma,\alpha (x+i\sigma y)\big] \qquad [\sigma = \pm 1]\:.
\end{align}
Indeed, plugging this choice in the first Maxwell equation, this becomes {(using $\sigma^2=1$)}:
\begin{align}\begin{split}
    (x+i\sigma y)[\alpha \omega -i\sigma\Omega (k_\sigma-i\alpha)]k_\sigma=0\:,
\end{split}\end{align}
which is satisfied for $k=\omega-\sigma\Omega$ and $\alpha=i\sigma \Omega$, confirming Eq.~\eqref{Eqn:Dispersion}. Similarly, the other three (spatial) Maxwell equations become {(again using $\sigma^2=1$)}:
\begin{align}\begin{split}
    \omega^2-k_\sigma^2+i\alpha k_\sigma-\omega\sigma\Omega&=-(x+i\sigma y)\left[\alpha \omega-(ik_\sigma+\alpha)\sigma\Omega\right]yk_\sigma\Omega\;,\\
    -i\sigma(\omega^2-k_\sigma^2+i\alpha k_\sigma-\omega\sigma\Omega)&=-(x+i\sigma y)\left[\alpha \omega-(ik_\sigma+\alpha)\sigma\Omega\right]xk_\sigma\Omega\;,\\
    (x+i\sigma y)[\alpha(\omega-\sigma\Omega)^2&-i\sigma\Omega k_\sigma(\omega-\sigma\Omega)]=0\:.
\end{split}\end{align}
One can easily verify that these equations are solved as well by the choice $k=\omega-\sigma\Omega$, $\alpha=i\sigma \Omega$.

\section{Generic rotations}
\label{app:generalRotations}
Our purpose in this appendix is to demonstrate that the phase shift of the transverse vector potentials considered in the main text {(and derived in detail in appendix \ref{app:rotatingframe})} still holds even when there are additional small rotations about axes orthogonal to the cavity symmetry axis.

Let the cavity symmetry axis be the $z$-axis. 
For a generic rotation $\vec{\Omega}\equiv(\Omega_x,\Omega_y,\Omega_z)$, the metric in the cavity-fixed frame would read:
\begin{align}
	g_{\mu\nu}dx^\mu dx^\nu=(-1+|\vec{\Omega}\wedge\vec{r}|^2)dt^2+2(\vec{\Omega}\wedge\vec{r})\!\cdot\! d\vec{r}\,dt+|d\vec{r}|^2\;.
\end{align}
Working in Weyl gauge, where the potential can be written as
\begin{align}
	A_{\mu} = \left(0, A_x,A_y,A_z \right) \equiv (0,\vec{A})\;,
\end{align}
we can perturb about the solution at \eqref{eq:simplifiedSoln}, which is exact when $\Omega_x=\Omega_y=0$.
Suppose that we treat $\Omega_x,\Omega_y$ as formal small parameters by sending $\Omega_{x,y} \rightarrow \epsilon \Omega_{x,y}$, and that we write each component of the vector potential as a power series in $\epsilon$.
Substituting into Maxwell's equations (MEs), expanding in powers of $\epsilon$, and satisfying the resulting equations that occur at each independent power of $\epsilon$ separately, allows us to derive a systematic perturbative expansion in powers of the formal small parameter $\epsilon$.
Moreover, we will also consider the solution only close to the symmetry axis of the cavity, such that we can take $x,y\rightarrow 0$ \emph{after} substituting into MEs.%
\footnote{Of course, because MEs are differential equations, it is important for the purposes of checking if this is a good solution to keep terms $\sim x^n,y^m$ until after substitution in the MEs. That is, to only set $x,y\rightarrow 0$ after such substitution.} %

The following perturbative expansion around the solution at \eqref{eq:simplifiedSoln} results from this procedure:
\begin{subequations}
\begin{align}
    A_\mu &= ( 0 , 1 , i \sigma , A_z ) e^{i\omega t - i k z} ; \quad k = \omega - \sigma \Omega_z \\
    A_z &= i \sigma \Omega_z (x+i\sigma y) + \sigma \frac{\Omega_x+i\sigma \Omega_y}{\omega-\sigma \Omega_z} \left[ 1 - i (\omega - \sigma \Omega_z) z - \frac{1}{4} \omega (2\omega -\sigma \Omega_z ) (x^2+y^2) \right]\:.
\end{align}
\end{subequations}
For the avoidance of doubt, this solution is approximate in the following sense. 
If $\Omega_x=\Omega_y=0$, this solution is exact. 
If $\Omega_x,\Omega_y\neq 0$, then after substituting into MEs, this solution fails to satisfy MEs by terms that fall into the following three categories:
(1) terms linear in $\Omega_i \ (i=x,y)$ [i.e., $\mathcal{O}(\epsilon)$] but which vanish at $x=y=0$; (2) terms quadratic in $\Omega_i\Omega_j \ (i=x,y)$ [i.e., $\mathcal{O}(\epsilon^2)$] that vanish at $x=y=0$; and (3) terms quadratic in $\Omega_i\Omega_j \ (i=x,y)$ [i.e., $\mathcal{O}(\epsilon^2)$] that do not vanish at $x=y=0$.

Let us spell out the orders of these terms more explicitly.
MEs are a series of second order differential equations as far as the vector potentials are concerned. 
The leading terms that cancel out upon substitution of the correct dispersion relation are of $\mathcal{O}(\omega^2,\omega k , k^2) \sim \mathcal{O}(\omega^2)$.
To make dimensionless comparisons with these overall leading terms in MEs, we will therefore imagine having divided MEs by $\omega^2$ after substitution, and will quote the orders of the \emph{leading} remaining dimensionless corrections at $\mathcal{O}(\epsilon,\epsilon^2)$ that fall into each of the three categories listed above, assuming that $\Omega_{x,y,z}\ll \omega$.
Category (1): terms of $\mathcal{O}(\omega x_i^2 \Omega_j ,\ \omega x_i^2x_j \Omega_k \Omega_z ,\ \Omega_i x_j )$ where $i,j,k=x,y$. 
Category (2): terms of $\mathcal{O}(\Omega_i\Omega_j x_k z,\ \Omega_i\Omega_j x_k / \omega ,\ \omega \Omega_i \Omega_j x_k x_\ell z,\ \Omega_i \Omega_j x_k x_\ell )$ where $i,j,k,\ell=x,y$.
Category (3): terms of $\mathcal{O}(\Omega_i^2z/\omega,\ \Omega_i\Omega_j/\omega^2)$ where $i,j=x,y$.

Therefore, so long as $x_i\Omega_j\ll 1$, $\Omega_z z \ll 1$, $\Omega_z x_i \ll 1$, $\Omega_i z \ll 1$, these terms are all automatically small, with the possible exception of terms of $\mathcal{O}(\omega x_i^2\Omega_j)$ [category (1)] and those of $\mathcal{O}(\omega x_kx_\ell z \Omega_i \Omega_j)$ [category (2)], which contain factors $\sim \omega z , \omega x_i$, which are in general large. 
To guarantee that these terms are also small, it suffices to impose one more requirement: $\omega \Omega_i x_j x_k \ll 1$ (note: this is parametrically sufficient, but not necessary if the previous conditions are strongly numerically satisfied). 
Roughly, and up to numerical factors, this says that the transverse beam size should be parametrically smaller than the geometric mean of the EM wavelength and the inverse rotation rate about the $x,y$ axes. 

Under these conditions, which are all easy to satisfy, we see also that the $A_z$ terms are also small corrections.
Moreover, the transverse vector-field components $A_x,A_y$ are not corrected at $\mathcal{O}(\epsilon)$, and nor is the leading dispersion relation.

We therefore conclude that, as long as the mild conditions derived in this appendix hold, the approach we advanced in the main text will continue to apply even when $\Omega_{x,y} \neq 0$, so long as these rotation rates are sufficiently small.


\section{Inertial frame and rotation of the apparatus}
\label{app:Inertial}
	
We can describe the system in the inertial frame by solving the ordinary field equations $\partial_\mu F^{\mu\nu}=0$ for the inertial vector field $A_\mu$, provided appropriate boundary conditions that describe a rotating source.
In the case of a rotation $\Omega$ around the $\hat{z}$ axis, for a planar source of circularly polarized light localized at $z=0$ the field will take the form $A_\mu=\epsilon_\mu e^{i(\omega_{*} t-kz)}$. At the boundary $z=0$ the field will take the form $ A^B_\mu(t)=e^{i\omega t}(0,e^{-i\sigma\Omega t},i\sigma e^{-i\sigma\Omega t},0)$ and the boundary conditions will read $A_\mu(t,z=0)=A^B_\mu(t)$. This implies $ \omega_*=\omega-\sigma\Omega$ so that the solution to the inertial equations of motion is:
\begin{align}
	A_\mu=e^{i[(\omega-\sigma\Omega )t-kz]}(0,1,i\sigma,0)\quad,\quad k=\omega-\sigma\Omega\;.
\end{align}
This solution also satisfies the condition $A_\mu(t,z)=A_\mu^B(t-\frac{k}{\omega_*}z)$, meaning that as the wavefront propagates, the polarization stays the same as it was at the moment of emission. The argument above shows that for circular polarization, regardless of the ratio $\Omega/\omega$, the resulting field can be expressed in terms of a single circularly polarized mode.
    
After the field has traveled for an interval $L$ in the $\hat{z}$ direction, two signals with opposite circular polarizations can be interfered to read the phase shift. This happens through a rotating apparatus, for which the basis of circularly polarized light is given by $\epsilon_\sigma=(0,e^{-i\sigma\Omega t},i\sigma e^{-i\sigma\Omega t},0)$. In this basis, the signal $A_\mu(t,L)$ will still have a polarization-dependent phase:
\begin{align}
	A_\mu= \epsilon_\sigma e^{i(\omega t-kL)}=\epsilon_\sigma e^{i\omega (t-L)}e^{i\sigma\Omega L}\;.
\end{align}
This makes clear that reading the signal from a rotating laboratory does not remove the polarization-dependent phase shift. Indeed, while the time-dependent part of the phase shift between different polarizations is absorbed by the basis, the space-dependent relative phase accumulated over the interval $L$ does not cancel. This is the same phase shift one sees in the non-inertial frame.

If the signal emitted was, e.g., linearly polarized in the rotating frame, then the situation would be different. In the inertial frame the boundary condition would be: $A_\mu^B(t)=e^{i\omega t}(0,\cos (\Omega t), -\sin (\Omega t),0)$, meaning that in this case the propagating field would be expressed as a linear combination of circularly polarized modes:
\begin{align}
	2A_\mu=e^{i[(\omega-\Omega )t-k_+z]}(0,1,i,0)+e^{i[(\omega+\Omega )t-k_-z]}(0,1,-i,0)\quad,\quad k_\pm=\omega\mp\Omega\;.
\end{align}
This solution does again satisfy $A_\mu(t,z)=A_\mu^B(t-z)$, meaning that as a wavefront propagates it remains a linear combination of the two circular polarizations emitted initially. 
Since these have different dispersion relations, they accumulate a relative phase, resulting in a rotation of the plane of linear polarization (cf.~\cite{PhysRevD.41.1231,PhysRevD.43.3789,HARARI199267,PhysRevLett.81.3067,DeRocco:2018jwe,Liu:2018icu,Fedderke:2019ajk}).
\bibliographystyle{JHEP}
\bibliography{references}

\end{document}